\documentclass[10pt,twocolumn,preprintnumbers,amsmath,amssymb,nofootinbib,superscriptaddress]{revtex4-1}

\usepackage{epsfig}
\usepackage{url}
\usepackage{hyperref}

\usepackage[normalem]{ulem}

\usepackage{latexsym}
\usepackage{epsfig}
\usepackage{amsmath}
\usepackage{amssymb}
\usepackage{wasysym}
\usepackage{graphicx}
\usepackage{verbatim}
\usepackage{enumerate,mdwlist}
\usepackage[titletoc]{appendix}
\usepackage{amsfonts}
\usepackage{pgfplots}
\usepackage[export]{adjustbox}
\usepackage{bbold}
\usepackage[normalem]{ulem}

\newcommand{\lp}{\left(}
\newcommand{\rp}{\right)}

\newcommand{\ba}{\begin{eqnarray}}
\newcommand{\ea}{\end{eqnarray}}
\newcommand{\be}{\begin{equation}}
\newcommand{\ee}{\end{equation}}

\newcommand{\dd}{\mathrm{d}}
\newcommand{\yr}{\mathrm{yr}}
\newcommand{\obs}{\mathrm{obs}}
\newcommand{\mmax}{m_\mathrm{max}}
\newcommand{\mmin}{m_\mathrm{min}}
\newcommand{\gmax}{\gamma_\mathrm{max}}
\newcommand{\gmin}{\gamma_\mathrm{min}}
\newcommand{\detc}{\mathrm{det}}
\newcommand{\bbh}{\mathrm{cbc}}
\newcommand{\Gpc}{\mathrm{Gpc}}

\newcommand{\Msun}{M_\odot}

\newcommand{\R}{\mathcal{R}}

\newcommand{\Om}{\Omega_{\mathrm{m}}}
\newcommand{\siren}{spectral siren }
\newcommand{\sirenf}{``spectral siren'' }

\definecolor{grey}{rgb}{0.4,0.4,0.4}
\definecolor{dullmagenta}{rgb}{0.4,0,0.4}
\definecolor{darkblue}{rgb}{0,0,0.4}
\definecolor{midblue}{rgb}{0,0,0.5}
\definecolor{midred}{rgb}{0.5,0,0}
\definecolor{orange}{rgb}{1,0.5,0}
\definecolor{lightbrown}{rgb}{0.75,0.5,0.25}
\definecolor{tan}{cmyk}{0.14,0.42,0.56,0}
\definecolor{djunglegreen}{cmyk}{0.99,0,0.52,0}
\definecolor{lightgreen}{rgb}{0,1,0}
\definecolor{olivegreen}{cmyk}{0.64,0,0.95,0.40}
\definecolor{midgreen}{rgb}{0.0,0.675,0.0}
\definecolor{darkgreen}{rgb}{0,0.5,0}

\usepackage[all]{xy} 
\usepackage{amsfonts}

\begin{document}

\title{Spectral Sirens: Cosmology from the Full Mass Distribution\\of Compact Binaries}

\author{Jose Mar\'ia Ezquiaga}
\email{NASA Einstein fellow; ezquiaga@uchicago.edu}
\affiliation{Kavli Institute for Cosmological Physics and Enrico Fermi Institute, The University of Chicago, Chicago, IL 60637, USA}

\author{Daniel E. Holz}
\affiliation{Kavli Institute for Cosmological Physics and Enrico Fermi Institute, The University of Chicago, Chicago, IL 60637, USA}
\affiliation{Department of Physics, Department of Astronomy \& Astrophysics, The University of Chicago, Chicago, IL 60637, USA}

\begin{abstract}
We explore the use of the mass spectrum of neutron stars and black holes in gravitational-wave compact binary sources as a cosmological probe. These standard siren sources provide direct measurements of luminosity distance. In addition, features in the mass distribution, such as mass gaps or peaks, will redshift, and thus provide independent constraints on their redshift distribution. We argue that the entire mass spectrum should be utilized to provide cosmological constraints. For example, we find that the mass spectrum of LIGO--Virgo--KAGRA events introduces at least five independent mass ``features'': the upper and lower edges of the pair instability supernova (PISN) gap, the upper and lower edges of the neutron star--black hole gap, and the minimum neutron star mass. We find that although the PISN gap dominates the cosmological inference with current detectors (2G), as shown in previous work, it is the lower mass gap that will provide the most powerful constraints in the era of Cosmic Explorer and Einstein Telescope (3G). By using the full mass distribution, we demonstrate that degeneracies between mass evolution and cosmological evolution can be broken, unless an astrophysical conspiracy shifts all features of the full mass distribution simultaneously following the (non-trivial) Hubble diagram evolution. We find that this self-calibrating ``spectral siren'' method has the potential to provide precision constraints of both cosmology and the evolution of the mass distribution, with 2G achieving better than $10\%$ precision on $H(z)$ at $z\lesssim1$ within a year, and 3G reaching $\lesssim1\%$ at $z\gtrsim2$ within one month. 
\end{abstract}

\date{\today}

\maketitle

The expansion rate, $H(z)$, is a fundamental observable in cosmology. There has been intense focus on its local value, $H_0$, due to existing $\sim4\sigma$ tensions between some early and late Universe probes~\cite{Freedman:2017yms,Verde:2019ivm}. The full redshift distribution of $H(z)$ is also of great interest, since it is a direct probe of $\Lambda$CDM, and may help unveil the nature of dark energy and test general relativity (GR)~\cite{Frieman:2008sn,Clifton:2011jh,Ezquiaga:2018btd}. 
Compact binary coalescences can be used as standard sirens~\cite{Schutz:1986gp,Holz:2005df}: {from} the amplitude and frequency evolution of their gravitational-wave (GW) emission one can directly infer the luminosity distance to the source. This is a particularly powerful probe since it directly measures distance at cosmological scales without any sort of distance ladder, and the sources are calibrated directly by GR.
When complemented with electromagnetic counterparts, such as transient events or associated galaxy catalogs, one can infer the redshift and directly constrain cosmological parameters.
These bright and dark siren methods have been applied by the LIGO--Virgo Collaborations~\cite{Aasi2015,Acernese_2014,LIGOScientific:2017adf,LIGOScientific:2018gmd,DES:2019ccw,LIGOScientific:2020zkf,LIGOScientific:2019zcs,Virgo:2021bbr,DES:2020nay}, as well as by independent groups~\cite{Vasylyev:2020hgb,Finke:2021aom,Palmese:2021mjm,Gray:2021sew}. 
Cross-correlations of GWs and galaxy surveys may also constrain $H(z)$~\cite{Oguri:2016dgk,Mukherjee:2020hyn,Diaz:2021pem}.

Even in the absence of electromagnetic observations, GWs \emph{alone} can probe $H(z)$ if they are analyzed in conjunction with known astrophysical properties of the population of compact binaries. 
Cosmology fixes the way in which the observed (redshifted) masses scale with luminosity distance. Therefore, by tracking the mass spectrum in different luminosity distance bins one can infer the redshift of the binaries, transforming them into powerful standard sirens. 
This \sirenf method works best when there is a distinct and easily identifiable feature. 
Binary neutron stars (BNSs) were the first to be proposed 
due to the expected 
maximum upper limit on the mass of neutron stars~\cite{Chernoff:1993th,Taylor:2011fs}. 
The masses of binary black holes (BBHs) also show interesting features, including a pronounced dearth of BBHs at high mass~\cite{Fishbach:2017zga, LIGOScientific:2021psn}. This feature is thought to come from the theory of pair instability supernova (PISN)~\cite{Barkat:1967zz,Fowler:1964zz,Heger:2001cd,Fryer_2001,Heger_2003,2016A&A...594A..97B}, which robustly predicts a gap 
between $\sim50$--$120\Msun$. 
The lower edge of the PISN gap is a clear target for second-generation (2G) detectors~\cite{Farr:2019twy} and has been explored for third-generation (3G) interferometers~\cite{You:2020wju}. 
Constraints on $H_0$ from the latest 
catalog are $\sim 20\%$ at $1\sigma$~\cite{Virgo:2021bbr}. 
Second-generation detectors at A+ sensitivity and 3G could also detect far-side black holes on the other side of the gap, thereby resolving the upper edge of the PISN gap and providing another anchor for cosmography~\cite{Ezquiaga:2020tns}. 

We explore the capabilities of current and next-generation detectors to probe $H(z)$ with the \emph{full} mass distribution of compact binaries. 
Uncertainties in the astrophysical modeling of the mass spectrum~\cite{Mastrogiovanni:2021wsd,Mukherjee:2021rtw} 
can impact the cosmological inference. 
We focus on the possible biases induced by the evolution of the masses  
and demonstrate that these degeneracies can be broken with spectral sirens. By using the entire mass distribution, the population itself allows one to constrain potential systematics due to evolution---in this sense, spectral sirens are {\em self-calibrating}.
We concentrate on flat-$\Lambda$CDM, but our methods can be straightforwardly generalized to other models. 
Our method can also be used to test GR~\cite{Ezquiaga:2021ayr}, as demonstrated with current BBH data~\cite{Ezquiaga:2021ayr,Mancarella:2021ecn,Leyde:2022orh}, and could be extended  
to BNSs~\cite{Ye:2021klk,Finke:2021eio}. 
Measuring tidal effects in BNSs will provide additional redshift information given the universality of the equation of state of matter at nuclear density \cite{Messenger:2011gi,2021PhRvD.104h3528C}.


\textbf{\emph{Five independent probes of cosmology.}} 
Our understanding of the population of stellar-origin compact binaries is far from complete. However, current GWs catalogs already provide suggestive and interesting insights~\cite{LIGOScientific:2018jsj,LIGOScientific:2020kqk,LIGOScientific:2021psn}. 
In this Letter we focus on the mass distribution, for which a number of broad properties are already well constrained:
\begin{enumerate}[i)]
    \item A drop in the BBH rate above $\sim45\Msun$. This dearth of mergers of more massive BBHs is statistically robust, and coincides with the range of masses where LIGO--Virgo are most sensitive~\cite{Fishbach:2017zga,Ezquiaga:2020tns,LIGOScientific:2021psn}. Data suggest that this feature can be modeled with a broken-power law. 
    \item A drop in the rate 
    at $\sim2.5\Msun$ and a break at $\sim5\Msun$ in the power law behavior above this mass~\cite{Farah:2021qom,LIGOScientific:2021psn}.
    The sharp feature at $\sim2.5\Msun$ is statistically well resolved and robust, but data are inconclusive  
    as to the distribution within the putative gap at $\sim2.5$--$5\Msun$. 
    Overall, the most likely local rate of binaries with component masses below $\sim2.5\Msun$ is about 10 times larger than the rate above $\sim5\Msun$, although uncertainties are still large~\cite{LIGOScientific:2021psn}. 
\end{enumerate}
Interestingly, the evidence of i) is roughly consistent with the 
prediction for a  
\emph{PISN gap} or \emph{upper mass gap}. Since current sensitivities drop above the upper edge of the PISN gap, we are still agnostic about a possible population of \emph{far-side binaries} above this feature~\cite{Ezquiaga:2020tns}, although we have upper bounds on their rate~\cite{LIGOScientific:2021tfm}. 
On the other hand, ii) would be consistent with electromagnetic observations suggesting a \emph{NSBH gap} or \emph{lower mass gap}~\cite{Bailyn:1997xt,Ozel:2010su,Farr:2010tu}. 
GW data robustly suggest that both BNSs and BBHs cannot be described by a single power law, but it cannot conclusively resolve the precise nature of the gap~\cite{Farah:2021qom}. 
Sub-solar mass astrophysical binaries are currently disfavored by theory, since objects more compact than white dwarfs are not expected as the endpoint of stellar evolution in this mass range~\cite{Chandrasekhar:1931ih}. Furthermore, they are disfavored by data, as targeted searches have found no candidates~\cite{LIGOScientific:2021job}.

The evidence for these features in the mass distribution of compact binaries suggests that there will be at least five independent mass scales: 
the edges of the lower and upper mass gaps, as well as the minimum neutron star mass. 
Each of these scales can be used to anchor the mass distribution in the {\em source}\/ frame, and thus the {\em detector}-frame distributions will allow us to infer the redshift:
\begin{equation}
    z(d_L)=m^\text{det}_\text{edge}(d_L)/m_\text{edge} - 1\,,
\end{equation}
where $m_\text{edge}=m^\text{det}_\text{edge}(d_L=0)$. 
Our fiducial, toy-model population 
is composed of a uniform distribution of BNSs between 1 and 2.5$\Msun$, a broken-power law model for BBHs below the PISN gap between a minimum and maximum mass, and a uniform distribution of far-side binaries. 
The local rates are fixed to 
$\R_0^\text{bns}=320\,\Gpc^{-3}\yr^{-1}$, $\R_0^\text{bbh}=30\,\Gpc^{-3}\yr^{-1}$, and $\R_0^\text{far-side}=0.1\,\Gpc^{-3}\yr^{-1}$, respectively, being consistent with population analyses~\cite{LIGOScientific:2021psn} and upper limits on intermediate-mass black holes~\cite{LIGOScientific:2021tfm}---see Supplemental Material for technical details, which includes Refs. \cite{alex_nitz_2019_3546372,PhysRevD.93.044006,Mandel:2018mve,Farr:2019rap,Callister:2020arv,emcee,ChainConsumer}. 
Although 2G instruments detect a greater fraction of high mass sources due to selection effects, 3G instruments are expected to detect {\em all}\/ sources, and will be equally sensitive to BNSs and BBHs across the mass spectrum. 
We assume the merger rates follow the star formation rate \cite{Madau:2014bja}.

The real distribution of compact binaries will certainly be more complex than the above description. 
Additional features will be beneficial, since these will introduce extra reference scales that can be tracked in the same way as the edges of the mass gaps, for example, the current excess of detections at $\sim35\Msun$~\cite{LIGOScientific:2021psn}. 
Moreover, since edges are easier to find than peaks, the cosmological inference will be dominated by the gaps. 
The utility of these features will be related to their prominence, such as the sharpness of the edges.  
For simplicity, we consider the gap edges to be step functions---more detailed calculations with smooth transitions provide constraints at the same order of magnitude, see e.g.~\cite{Farr:2019twy}. 

\textbf{\emph{The lower mass gap will win.}} 
The constraints on $H(z)$ 
are most sensitive to how well we resolve the edges of the mass gaps, 
which is directly related to the number of events at these scales at different redshifts. 
These numbers will be a combination of the detector sensitivity and the intrinsic merger rate $\R(z)$. 
Despite the larger intrinsic rate of low-mass binaries, the selection effects of 2G detectors significantly reduces the detectability. This changes with 3G detectors \cite{Vitale:2016aso}, where essentially all astrophysical stellar-origin binaries are detected across cosmic history. 
Quantitatively, for 2G detectors we find $\sim6\%$ of detections having masses below $7\Msun$, and $\sim10\%$ above $45\Msun$. 
With 3G sensitivities these numbers shift to $\sim27\%$ below $7\Msun$ and $\sim2\%$ above $45\Msun$. 
This suggests that the lower mass gap will play an increasingly important role transitioning to 3G.

The precision in $H(z)$ depends on the errors in distance and redshift. The error in $d_L$ scales as $1/\sqrt{N}$, where $N$ is the relevant number of binaries providing the measurement (see e.g.~\cite{2006PhRvD..74f3006D,2018Natur.562..545C}). The error on the redshift is dominated by the uncertainty in locating features in the observed mass distribution. For example, the error in locating the ``edge'' of a mass gap is expected to scale as $1/N$~\cite{Ezquiaga:2020tns}, so long as the errors in the individual mass measurements are sub-dominant. Since, generally, distance is measured more poorly than mass, we find:
\begin{equation}
    \frac{\Delta H(z)}{H(z)}\sim \frac{\Delta d_L/d_L}{\sqrt{N_\text{edge}}}\,,
\end{equation}
where $N_\text{edge}$ is the number of events with information about the edge. 
We follow \cite{Fishbach:2019ckx} to simulate GW detections including selection effects and detector uncertainties.
Dividing the number of detections into four redshift bins for $z<2$, we estimate $\Delta H(z)/H(z)$ from each edge of the mass gaps. 

Although the lower edge of the PISN mass gap will dominate the inference of $H(z)$ with current detectors (reaching 5$\%$--10$\%$, in agreement with~\cite{Farr:2019twy}), it is the lower mass gap that will dominate the 3G inference, potentially reaching sub-percent precision. Moreover, with 3G detectors the precision in $H(z)$ is sustained beyond $z>1$.

\textbf{\emph{Degeneracies between cosmology and mass evolution can be broken.}} 
For spectral sirens, 
it is critical to understand if the mass distribution itself evolves, since such evolution might bias the inference of $H(z)=H_0\sqrt{\Om(1+z)^3 + 1 - \Om}$, where $H_0$ and $\Om$ are the expansion rate and (dimensionless) matter density today. 
For convenience we also introduce 
$h_0=H_0/(100\mathrm{km/s/Mpc})$.

In the context of 3G, and assuming $\R(z)$ peaks around $z\sim2$, the vast majority of binaries will be detected for all viable ranges of cosmological parameters~\cite{Planck:2018vyg}. 
We can therefore neglect any mass or redshift dependence in the detection probability, and the effect of modifying cosmology becomes transparent: 
a) it changes the overall rate as a function of redshift, and b) it shifts the detector frame masses of the entire population. 
For $\R(z)$ following the star formation rate, there is no clear correlation with $(H_0,\Om)$, except for the degeneracy between $\R_0$ and $H_0^3$.  
Importantly, the bulk of the cosmological constraints will come from the observed mass distribution rather than the overall rate.

\begin{figure}[t!]
\centering
\includegraphics[width = \columnwidth,valign=t]{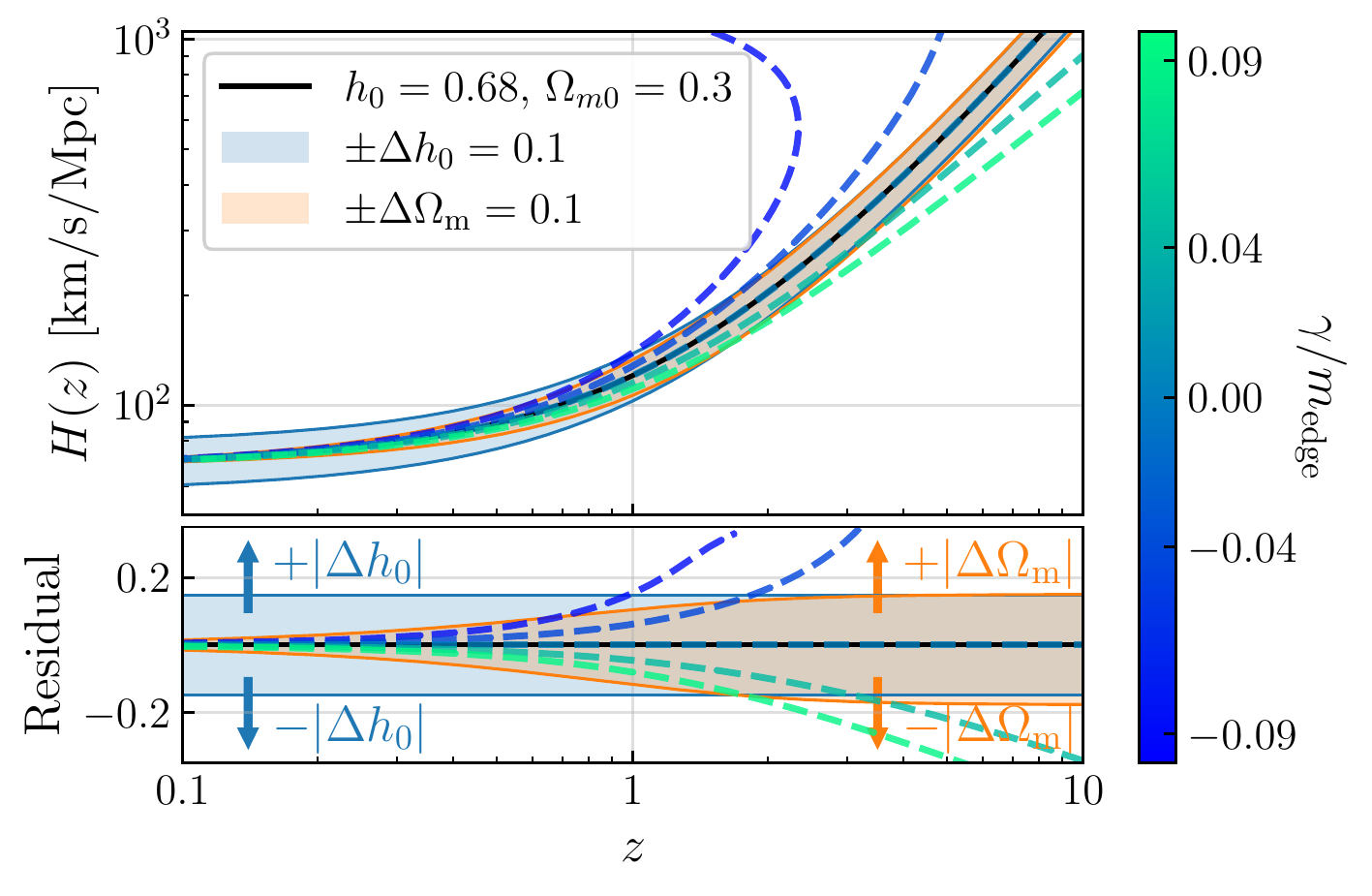}
\caption{Inferred Hubble parameter for different fiducial cosmologies (color bands) and evolutions of an edge of the mass distribution (dashed lines). We assume that $m_\mathrm{edge}$ is measured at low-redshift and the redshift is biased by the linear evolution $\gamma$, cf. (\ref{eq:z_bias}). The bottom panel corresponds to the residual against the fiducial cosmology: $(H(z)-H_\mathrm{fid}(z))/H_\mathrm{fid}(z)$, with $H_\mathrm{fid}(z)$ fixed by $h_0=0.68$ and $\Om = 0.3$ (black line). 
Arrows indicate the effect of changing $h_0$ and $\Om$.}
\label{fig:Hz_deg}
\end{figure}

The evolution of the mass distribution does not mimic the cosmology unless the \emph{entire}\/ spectrum shifts uniformly, so that the shape is completely unaltered. 
However, in general we would expect the evolution to change its shape, see e.g.~\cite{vanSon:2021zpk,Mapelli:2021gyv}, and, therefore, cosmology and evolution of the mass distribution can be disentangled. 
Nonetheless, we can imagine that time evolution might affect one of the edges of one of our mass bins. 
As an example, we consider a linear-in-redshift mass evolution controlled by $\gamma$, i.e. $m_\text{edge,ev}(z)= m_\text{edge}+\gamma\cdot z$, assuming that 
$m_\mathrm{edge}$ is measured at $z\sim0$. In this case the inferred redshift when not taking this evolution into account will be biased by
\begin{equation} \label{eq:z_bias}
    z_\text{bias}=(1+z)\lp 1+ \gamma\cdot z/m_\mathrm{edge}\rp -1\,.
\end{equation}
Consequently, if $\gamma>0$, $z_\text{bias}$ will be shifted toward higher values which, at fixed $d_L$, is equivalent to a larger $H_0$.  
For example, a $0.1\Msun$ shift of a $5\Msun$ edge at $z\sim1$ will change $H(z)$ by $\sim3\%$.   
Importantly, this is only an approximate degeneracy. As shown in Fig.~\ref{fig:Hz_deg}, when considering the Hubble diagram at all redshifts the effect of astrophysical evolution 
will not, in general, match with any allowable $\Lambda$CDM cosmologies.  
An evolution of the mass scale that mimics the low-$z$ effect of changing $h_0$ will overshoot the modification of $\Om$ at high-$z$. 
Because the larger differences occur at $z>1$, this figure helps us anticipate that 3G detectors will more effectively disentangle the astrophysical evolution 
from varying cosmology. 
Note that, although we have chosen a particular parametrization for $m_\text{edge,ev}(z)$, our conclusions  
hold, in general: we can disentangle evolution of the mass distribution so long as it is not perfectly tuned to change in accordance with the (highly nontrivial) Hubble diagram shown in Fig.~\ref{fig:Hz_deg} at all mass scales.

\textbf{\emph{Examples of 2G and 3G inference.}} 
To explore the degeneracy space between different cosmologies and astrophysical evolution, 
we generate a mock catalog of events and perform a Bayesian hierarchical analysis to infer $H(z)$. 
In particular, we study how the inference of $(h_0,\Om)$ is affected by the fiducial population model.  
We focus on BBHs between the lower and upper mass gaps, delimited by $\mmin$ and $\mmax$, considering 
three scenarios: 
\begin{enumerate}[1)]
    \item {\em no evolution}: the intrinsic population has fixed edges over cosmic time ($\gmin=\gmax=0\Msun$),
    \item {\em one-sided evolution}: the maximum mass increases with redshift ($\gmin=0\Msun$, $\gmax\neq0\Msun$),
    \item {\em independent two-sided evolution}: both the minimum and maximum masses evolve ($\gmin,\,\gmax\neq0\Msun$).
\end{enumerate}  
We consider 1,000 2G detections at A+ sensitivity~\cite{Aasi:2013wya,sensitivity_curves_ligo}, and 10,000 3G events at Cosmic Explorer sensitivity~\cite{sensitivity_curves_3g}. 
This corresponds to roughly 1 year and 1 month of observation, respectively. 
For simplicity, we restrict the Bayesian inference to the most relevant parameters: $\{h_0,\Om,\mmin,\mmax,\gmin,\gmax\}$; we have checked explicitly that this assumption does not affect our conclusions. 
Full posterior samples can be found in the Supplemental Material.

\begin{figure}[t!]
\centering
\includegraphics[width = \columnwidth,valign=t]{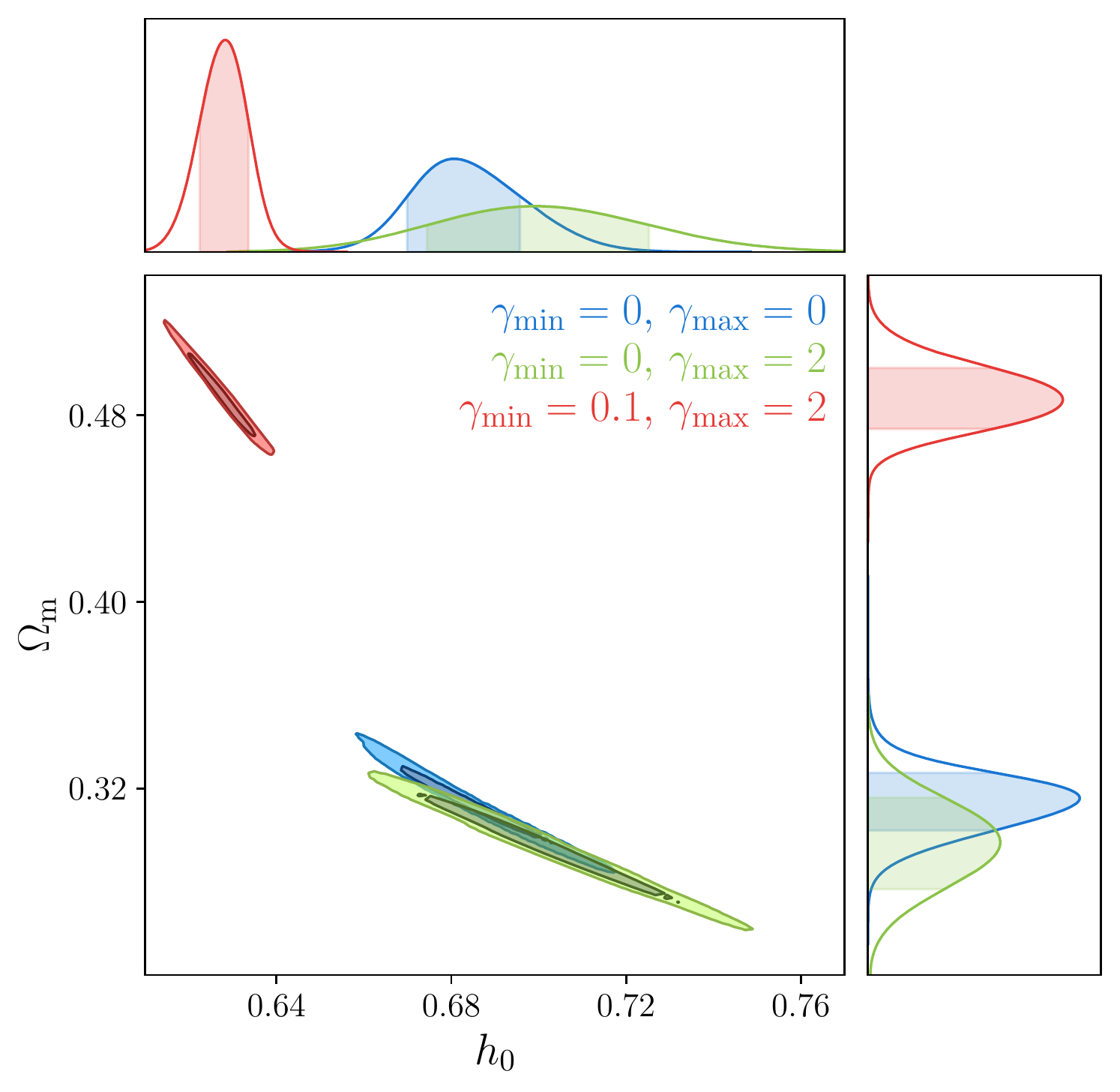}
\caption{Cosmological inference with 
10,000 3G BBH detections between fixed lower and upper mass gaps, when the fitting model does \emph{not} include evolution in the mass distribution. 
Different posterior distributions correspond to three mock populations: without evolution (blue), with evolution of the maximum mass (green), and evolution of both minimum and maximum masses (red). When not accounted for, the evolution of the mass distribution can bias $h_0$ and $\Om$. 
}
\label{fig:corner_plots_noev}
\end{figure}

We first analyze how the cosmological inference changes for different mock populations when the potential evolution of the mass distribution is not incorporated in the parameter estimation. 
When the mock catalog does not evolve, 
the fiducial cosmological parameters are well recovered since the fitting model matches the simulated data. 
However, when the catalogs include evolution, the inference of $(h_0,\Om)$ can become biased. 
This is especially acute for 3G, as plotted in  Fig.~\ref{fig:corner_plots_noev}. 
In this case, the larger bias occurs for the case of an evolving minimum mass ($\gmin\neq0$, red posteriors), since this is the scale controlling the inference. 
The red contours show that, since the minimum mass is increasing with redshift, the inferred redshifts are biased high, and thus to compensate $h_0$ is biased to lower values and $\Om$ is biased to higher values (see Fig.~\ref{fig:Hz_deg}). 
In the case where the minimum mass does not evolve while the maximum mass does evolve (green contours), the cosmology is not biased significantly, but the errors enlarge as the inference from the upper mass is degraded. 

We also analyze the same mock catalogs including $\gmin$ and $\gmax$ as free parameters. 
We find that $h_0$ and $\Om$ are no longer biased, although $\Om$ is now poorly constrained with 2G detectors in agreement with~\cite{Mastrogiovanni:2021wsd}, with the evolution parameters recovered at $\Delta\gmin/\gmin\sim20\%$ and $\Delta\gmax/\gmax\sim8\%$.  
3G places significantly better constraints with more accuracy at low masses, $\Delta\gmin/\gmin\sim1\%$ and $\Delta\gmax/\gmax\sim3\%$. 

\begin{figure}[t!]
\centering
\includegraphics[width = \columnwidth,valign=t]{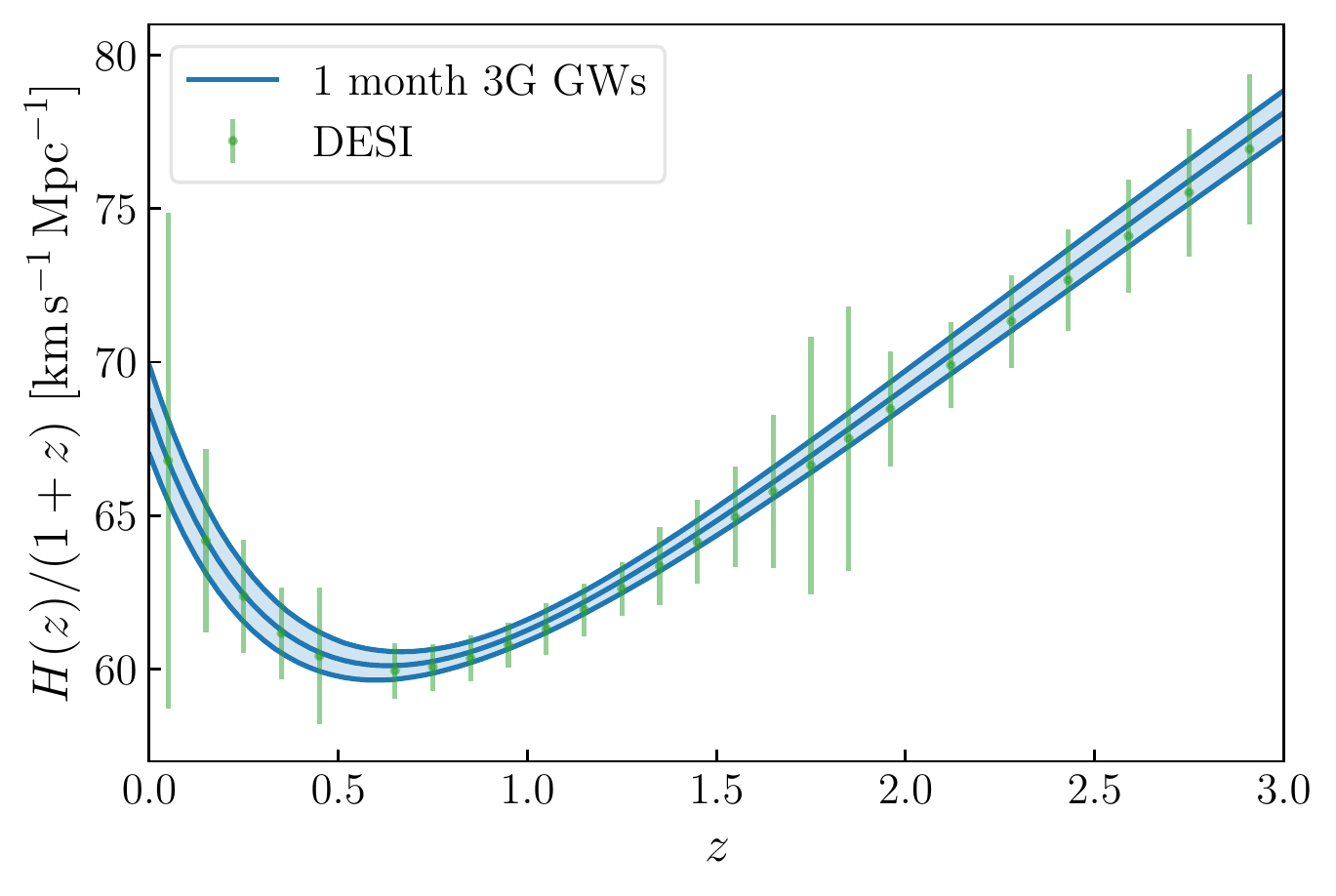}
\caption{Hubble parameter with 10,000 3G detections when the fitting model \emph{does} account for evolution in the mass distribution. 
Shaded region represents $1\sigma$ errors. 
Sub-percent precision is possible with 1 month of observation. We compare to DESI forecasts~\cite{DESI:2016fyo}.}
\label{fig:Hz}
\end{figure}

These results indicate that both 2G and 3G detectors will be able to simultaneously constrain cosmology and measure the evolution of the mass distribution. 
Translated into a measurement of the expansion rate (at $68\%$ C.L.), 2G detectors will within a year constrain $H(z)$ with better than $10\%$ accuracy at $z\sim 0.7$. 
This result is slightly inferior to the previous $\sim6\%$ forecast of $H(z\sim0.8)$ in one year from the PISN mass gap~\cite{Farr:2019twy}. This is because we adopt the latest fit to the data, consisting of a broken power-law with a less pronounced lower edge of the gap~\cite{LIGOScientific:2021psn}. A similar conclusion was found in~\cite{Mastrogiovanni:2021wsd}. 
Impressively, 3G detectors will within a month constrain $H(z)$ with $<1\%$ accuracy beyond $z\sim 1$, comparing favorably to measurements such as DESI~\cite{DESI:2016fyo} as shown in Fig.~\ref{fig:Hz}. 
Although our modeling of the mass distribution is only a rough parametrization, it provides a useful estimate of the capabilities of future detectors. 
We leave a detailed cosmological forecast with realistic BBH mass distributions from different formation channels for future work.  

\textbf{\emph{Future prospects.}} 
Next-generation GW detectors will perform sub-percent precision cosmography with standard sirens, providing a potentially revolutionary new cosmological probe.
A detailed understanding of the attendant systematics will be required to attain robust constraints.  
We have shown that in the 3G-era the specral siren measurement of $H(z)$ 
will be dominated by features associated with the lower mass gap. Moreover, by incorporating the possibility of redshift evolution of the intrinsic mass distribution, it is possible to simultaneously constrain such evolution along with the underlying cosmological model. 

We emphasize the utility of using the full mass distribution, rather than focusing on just one feature such as the lower edge of the upper or PISN mass gap. 
Each of the edges of the mass gaps (or any other relevant feature) can be thought of as providing an independent cosmological measurement. 
One can compare the values of $H_0$ and $\Om$ from each individual bump and wiggle and dip in the mass distribution, 
and in this way the GW population is {\em self-calibrating}: it can simultaneously constrain cosmology while testing for consistency and unearthing systematics due to population evolution. 
Alternatively, a Bayesian analysis of the entire catalog 
helps to narrow down the errors in $H(z)$ and simultaneously constrain the astrophysical evolution of the mass distribution. 
This work considers a simple, toy-model description for the mass distribution.  
In practice, the \siren method will utilize the full data-informed mass distribution incorporating all of the identifiable features  
simultaneously. 
The results can be thought of as a conservative estimate of the future potential of this approach and have implications both for cosmology and for our understanding of the formation and evolution of the relevant astrophysical populations.

One of the outstanding challenges of GW astrophysics is to develop an understanding of the formation channels that account for GW observations (see e.g.~\cite{2021ApJ...910..152Z,Mandel:2021smh}). 
Recent work has explored the evolution of the mass distribution in field binaries~\cite{vanSon:2021zpk} and clusters~\cite{Mapelli:2021gyv,2022arXiv220508549Z}, and searches have been performed in current data~\cite{Fishbach:2021yvy}. 
These works show that the high-mass end of the distribution is more susceptible to environmental effects, such as metallicity, that are expected to evolve with cosmic time, as well as the time delay distribution that affects the observed relative rates~\cite{vanSon:2021zpk,Mukherjee:2021rtw}. 
It is encouraging that current results indicate that the low-mass end of the spectrum could be more robust against redshift evolution, while 
providing the strongest constraints on $H(z)$. 
Given the potential scientific impact of the lower mass gap for cosmology, further exploration of its properties 
is warranted. 
Moreover, although other quantities such as mass ratios and spins do not redshift, if their intrinsic distributions evolve in time in a known fashion, they could provide redshift information as a spectral siren. 

Since spectral siren cosmology 
is a pure GW measurement, it is completely independent from results based on EM observations.
We have shown that the GW constraints compare favorably with current $\sim6\%$--$10\%$ baryon acoustic oscillations constraints from BOSS~\cite{duMasdesBourboux:2017mrl,Bautista:2017zgn,Zarrouk:2018vwy} and future forecasts, such as the $>2\%$ measurements expected from DESI~\cite{DESI:2016fyo} at $H(z>2)$. 
The spectral siren method is complementary to the bright siren approach, which uses EM counterparts to GW sources to constrain the redshift of the sources~\cite{Schutz:1986gp,Holz:2005df,2006PhRvD..74f3006D}. For ground-based detectors, the most promising counterpart sources are short gamma-ray bursts associated with BNSs. 
While these may be detectable to 
$z\gtrsim0.5$, they are likely inaccessible at $z\gtrsim1.5$~\cite{Belgacem:2019tbw,Chen:2020zoq}. 
Thus, spectral sirens will provide unique precision high-redshift constraints on both GW astrophysics and cosmology.

\begin{acknowledgments}
We are grateful to Amanda Farah, Will Farr, and Mike Zevin for stimulating conversations. 
We also thank Amanda Farah and Rachel Gray for comments on the draft, and Antonella Palmese  and Aaron Tohuvavohu for useful correspondence. 
JME is supported by NASA through the NASA Hubble Fellowship grant HST-HF2-51435.001-A awarded by the Space Telescope Science Institute, which is operated by the Association of Universities for Research in Astronomy, Inc., for NASA, under contract NAS5-26555. DEH is supported by NSF grants PHY-2006645 and PHY-2110507. DEH also gratefully acknowledges support from the Marion and Stuart Rice Award.
Both authors are also supported by the Kavli Institute for Cosmological Physics through an endowment from the Kavli Foundation and its founder Fred Kavli. 
\end{acknowledgments}

\appendix

\section*{Supplemental material}

\section{Methods}
\label{app:methods}

We summarize our methodology, detailing the observing scenarios, mock catalogs and Bayesian analysis.

\paragraph{Observing scenarios:} \label{app:observing_scenarios}

For second-generation detectors we use the A+ sensitivity curve of advanced LIGO described in~\cite{Aasi:2013wya}, which can be found at~\cite{sensitivity_curves_ligo} and is expected for O5 (2025+).
For third-generation detectors (2030+), we adopt the sensitivity curve of Cosmic Explorer given in~\cite{sensitivity_curves_3g}. We have checked that for Einstein Telescope the results are qualitatively the same.

\begin{figure*}[t!]
\centering
\includegraphics[width = 0.95\textwidth,valign=t]{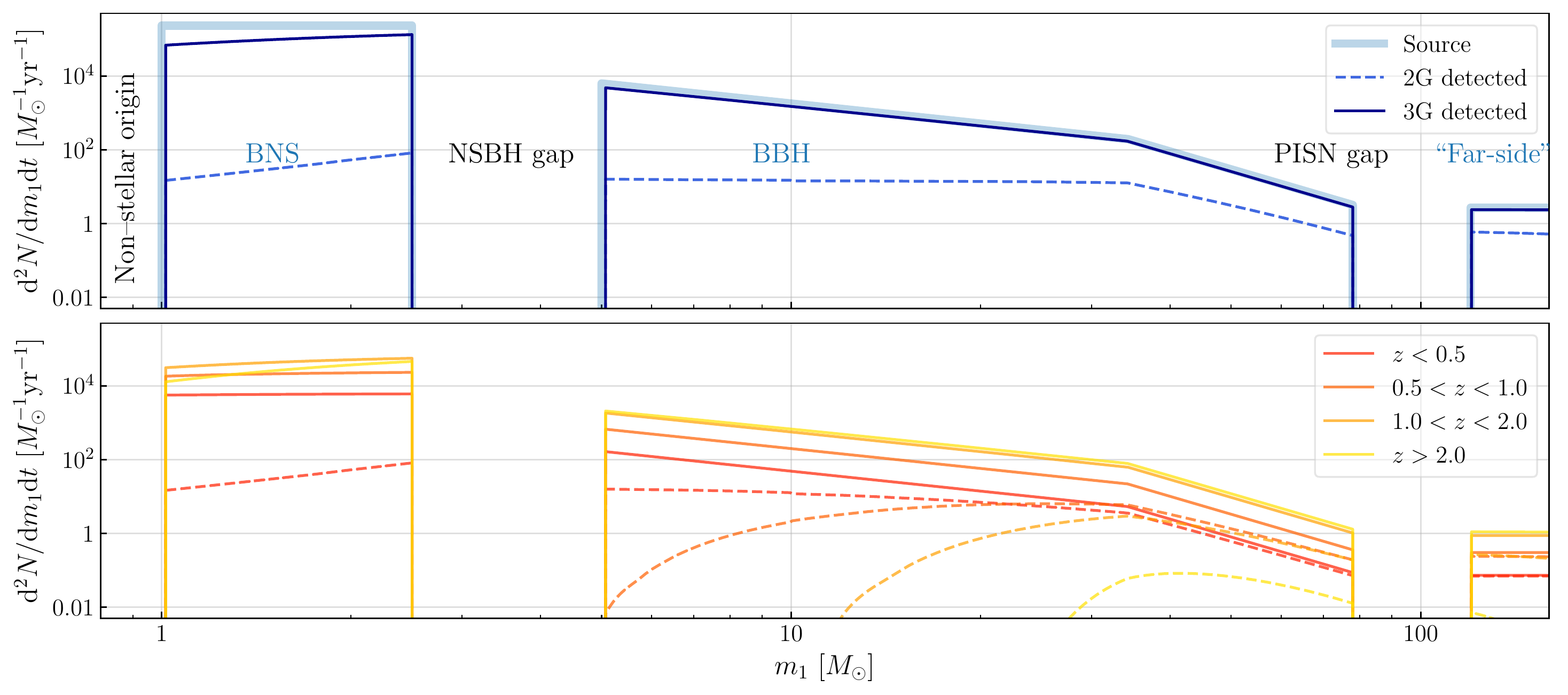}
 \caption{Toy model for the merger rate of compact-binary coalescences, as a function of the primary mass. The detected distribution of source-frame masses for second-detectors (2G) at A+ sensitivity~\cite{Aasi:2013wya} and third-generation (3G) detectors are plotted with thin-dashed and solid lines respectively. The intrinsic source mass distribution is presented with a thick line and it is characterized by two mass gaps: the neutron star black hole (NSBH) gap and the pair instability supernova (PISN) gap. This divides the population into binary neutron stars (BNSs), binary black holes (BBHs) and ``far-side" binaries in order of increasing mass. Note that NSBH binaries are formed from the BNS distribution and one of the others. The population of astrophysical binaries is closed at low masses by the minimum neutron star mass.
 The bottom panel shows the 2G and 3G detected distributions at different redshifts. 
 }
 \label{fig:dNdm1_all}
\end{figure*}

\paragraph{Mock catalog of GW detections:}

For a given compact binary population model we obtain the mock source parameters $m_1,\,m_2$ and $z$ drawing samples from their comoving merger rate $\R(z)$ and their source mass distribution $p(m_{1},m_{2}|z)$, which may depend on redshift (see App.~\ref{app:pop} for the specific models considered). 
In order to obtain the simulated detected posterior samples of $m_{1z},\, m_{2z}$ and $d_L$, we need to estimate the measurement errors and selection biases of a given GW detector. 
We use the methodology described in~\cite{Ezquiaga:2020tns}, which follows the prescription of~\cite{Fishbach:2019ckx}. 
We use \texttt{pyCBC}~\cite{alex_nitz_2019_3546372} with the \texttt{IMRPhenomD} approximant~\cite{PhysRevD.93.044006} to compute the signal-to-noise ratio (SNR) of a waveform of non-spinning compact binaries. 
We set the single-detector detection threshold at SNR of 8. 
We fix the fiducial cosmology to Planck 2018~\cite{Planck:2018vyg}: $H_0=67.66$km/s/Mpc and $\Om=0.31$.

\paragraph{Hierarchical Bayesian analysis}

The main output of the Bayesian inference is the posterior distribution of a given set of parameters $\Lambda$ describing a given population of compact binaries. This distribution follows from
\be
p(\Lambda|\{d_i\})\propto p(\{d_i\}|\Lambda)\pi(\Lambda)\,, 
\ee
where $p(\{d_i\}|\Lambda)$ 
is the likelihood of having $N_\obs$ events with data $\{d_i\}$, while $\pi(\Lambda)$ is the prior expectation on $\Lambda$. 

The likelihood of resolvable detections is described by a Poissonian process
\be
\begin{split}
p_\bbh(\{d_i\}|\Lambda)\propto &N_\detc(\Lambda)^{N_\obs}e^{-N_\detc(\Lambda)} \\
&\times\prod_{i=1}^{N_\obs}\frac{1}{\xi(\Lambda)}\left<\frac{p(\phi_i|\Lambda)}{\pi_\mathrm{pe}(\phi_i)}\right>_\mathrm{samples}\,,
\end{split}
\ee
where $\xi=N_\detc/N_\bbh$ is the ratio between the expected detected mergers $N_\detc$ and the actual merger $N_\bbh$. 
This quantity encodes all the selection effects. 
Note that the data likelihood, $p(d_i|\phi)$, given the GW parameters $\phi$, is not directly accessible. Instead there are only the event posteriors samples $p(\phi|d_i)$ to which it is necessary to factor out the prior used in the parameter estimation $\pi_\mathrm{pe}(\phi)$. 

Our main observables are the inferred redshift and source masses: $\phi=\{z,m_1,m_2\}$. These quantities depend on the cosmology $(h_0,\Om)$ and are derived from the observed data of $\{m_{1z},m_{2z},d_L\}$. 
The prior $\pi_\text{pe}(z,m_1,m_2)$ is directly obtained including the Jacobian since we use a flat prior in our mock simulations:
\be
\pi_\text{pe}(z,m_1,m_2)\propto (1+z)^2\frac{\partial d_L}{\partial z}\,,
\ee 
where in this case
\be
\frac{\partial d_L}{\partial z} = \frac{d_L}{(1+z)} +\frac{(1+z)d_H}{E(z)}\,, 
\ee 
with the luminosity distance defined as 
\begin{equation}
    d_L=(1+z)d_H\int_0^z\frac{\dd z'}{E(z')}\,,
\end{equation}
the horizon distance $d_H=c/H_0$ and $E(z)=H(z)/H_0$. In this notation the differential comoving volume is given by
\begin{equation}
    \frac{\dd V_c}{\dd z} = \frac{4\pi d_L^2 d_H}{(1+z)^2E(z)}\,.
\end{equation}
The total likelihood then reads 
\be
\begin{split}
p_\bbh(\{d_i\}|\Lambda)\propto &\,e^{-N_\detc(\Lambda)} \times\prod_{i=1}^{N_\obs}\left<\frac{dN(\phi|\Lambda)/d\phi}{\pi_\mathrm{pe}(\phi)}\right>_\mathrm{samples}\,.
\end{split}
\ee
Marginalizing the local merger rate $\R_0$ using a uniform in log prior, the above expression simplifies to
\be
\begin{split}
p_\bbh(\{d_i\}|\Lambda)\propto &\,\xi^{-N_\obs} \times\prod_{i=1}^{N_\obs}\left<\frac{p(\phi|\Lambda)}{\pi_\mathrm{pe}(\phi)}\right>_\mathrm{samples}\,,
\end{split}
\ee
which does not depend on $\R_0$. 
For a general discussion of this statistical framework see e.g.~\cite{Mandel:2018mve}. 

Finally let us note that the selection function necessary to compute $\xi$ is calculated by performing an injection campaign following~\cite{Farr:2019rap} and ensuring that the effective number of independent draws is at least 4 times larger than the number of detections in the catalog.

\begin{figure*}[t!]
\centering
\includegraphics[width = 0.95\textwidth,valign=t]{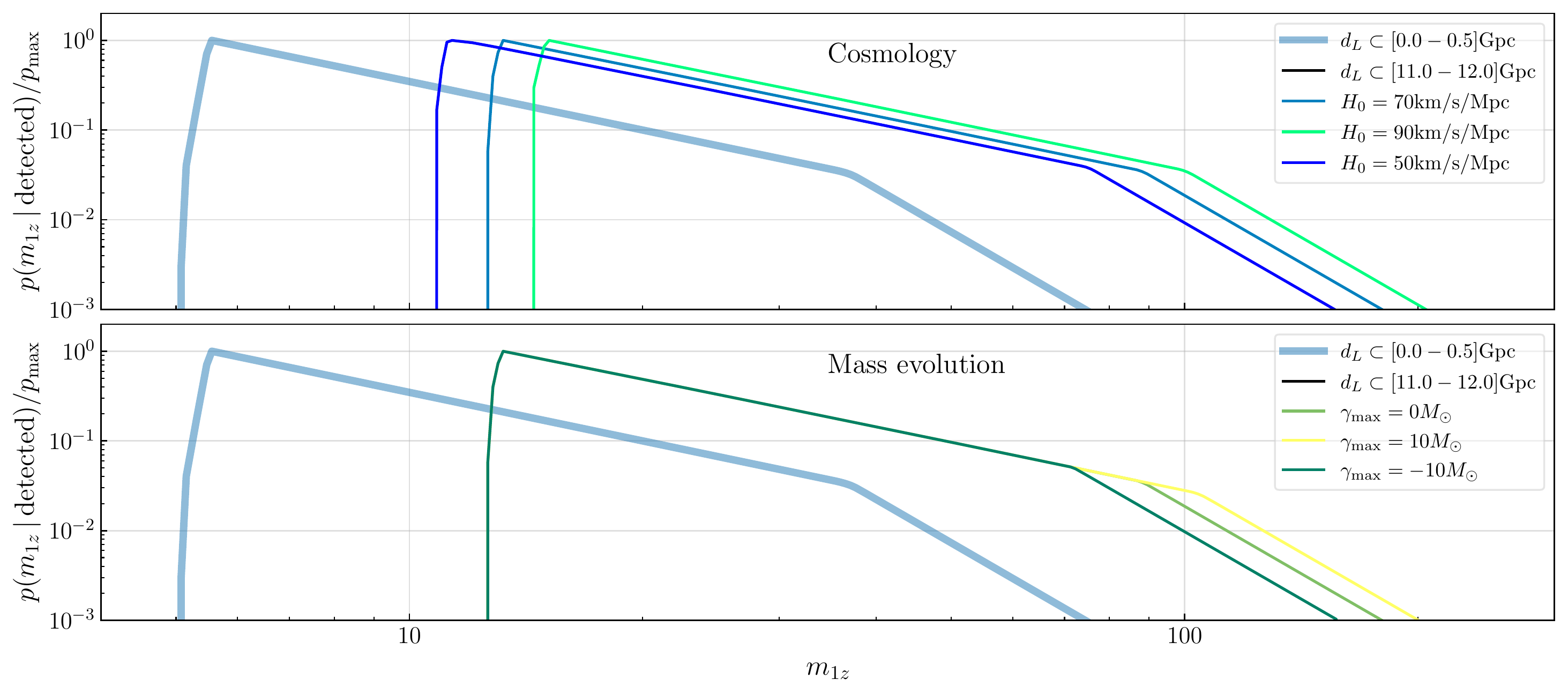}
\caption{Detected distribution of redshifted masses for a population of BBHs at different luminosity distance bins with 3G sensitivities. The BBHs population is modeled with an intrinsic broken power law source mass distribution between the lower and upper mass gaps. The top panel shows the effect of changing the cosmology, while the lower panel shows the effect of increasing the maximum mass with redshift following $m_\text{max,ev}=\mmax+\gmax\cdot z$. We normalize all the probability distributions w.r.t. their maximum.}
\label{fig:p_m1z}
\end{figure*}

\section{Full population of compact binaries}
\label{app:pop}

In order to fix the population of compact binaries we need to specify their comoving merger rate $\R(z)$ and their mass distribution $p(m_1,m_2|z)$. 
For all the mass spectrum we fix the merger rate with the same functional form inspired by the star formation rate~\cite{Callister:2020arv}:
\begin{equation}
    \R(z) \sim \R_0 \frac{(1+z)^\alpha}{1+\lp\frac{1+z}{1+z_p}\rp^{\alpha+\beta}}\,,
\end{equation}
with $\alpha = 1$, $\beta = 3.4$ and $z_p = 2.4$. The local merger rate is fixed to: 
$\R_0^\text{bns}=320^{+490}_{-240}\Gpc^{-3}\yr^{-1}$, $\R_0^\text{bbh}=30\pm15\Gpc^{-3}\yr^{-1}$ and $\R_0^\text{far-side}=0.1\Gpc^{-3}\yr^{-1}$. 

The BNS primary mass distribution is modeled as a uniform distribution between 1 and $2.5\Msun$. 
The BBH primary mass distribution is described a broken power-law:
\begin{equation}
p(m_1)\propto
\begin{cases}
m_1^{\kappa_1}, \quad \mmin<m_1<m_\text{break} \\
m_1^{\kappa_2}, \quad m_\text{break}<m_1<\mmax \\
0, \quad \text{elsewhere}
\end{cases}\,,
\end{equation}
where $m_\text{break} = \mmin + b(\mmax-\mmin)$ and $b\subset(0,1]$. In our fiducial model: $\mmin=5\Msun$, $\mmax=78\Msun$, $\kappa_1=1.8$, $\kappa_2=5$ and $b=0.4$.
For the far-side binaries we take a uniform distribution between 120 and $160\Msun$. 
In all the cases the secondary source mass is uniformly sampled between the minimum mass of the population and $m_1$. 
In the top panel of Fig.~\ref{fig:dNdm1_all} we show a sketch of our fiducial, toy-model population. This panel also  shows the expected detected distributions, incorporating selection effects. 
The rate of events in different redshift bins is shown in the lower panel

In order to model the possible mass evolution we shift the minimum or maximum mass or both linearly in redshift: $m_\text{edge,ev}=m_\text{edge}+\gamma\cdot z$. We denote $\gmin$ and $\gmax$ to the linear evolution affecting the minimum and maximum mass respectively. 
It is to be noted that we do not expect the local source frame astrophysics to know about the redshift and a more realistic choice would be to parametrize it with the logarithm of the metallicity. However, for a small mass evolution the linear relation considered here is a good approximation. 
We leave for future work a detailed study of parametrizations of the mass evolution from different formation channels. 

An example of the effect of this type of evolution of the mass distribution is presented in Fig.~\ref{fig:p_m1z} together with the comparison with the effect of changing the cosmology via $H_0$. 
In the detector frame, the cosmology just adds a constant shift to $\log m_{1z}$. Note that if one focuses on the maximum mass, around $100\Msun$, one could confuse the mass evolution with cosmology for a given luminosity distance bin. However, if we look at the full mass distribution, in particular the minimum mass around $10\Msun$, this degeneracy can be broken. This together with the different redshift evolution presented in Fig.~\ref{fig:Hz_deg} graphically explains why we can disentangle the cosmology from astrophysical evolution in our analysis.  

\section{Full posterior samples}
\label{app:posteriors}

We sample the posterior distribution using the MCMC code \texttt{emcee}~\cite{emcee}. We run the chains until convergence, ensuring that the effective number of samples, defined as the number of MCMC steps divided by the autocorrelation time, is at least larger than 100 for all the parameters. 
The priors are chosen to be uniform distributions in the ranges: $h_0\subset[0.2,1.2]$, $\Om\subset[0.1,0.9]$, $\mmin\subset[1,20]\Msun$, $\mmax\subset[30,150]\Msun$ and $\gmin,\gmax\subset[-10,10]\Msun$. 
The posterior samples are plotted using \texttt{ChainConsumer}~\cite{ChainConsumer}. 

\begin{figure*}[t!]
\centering
\includegraphics[width = 0.75\textwidth,valign=t]{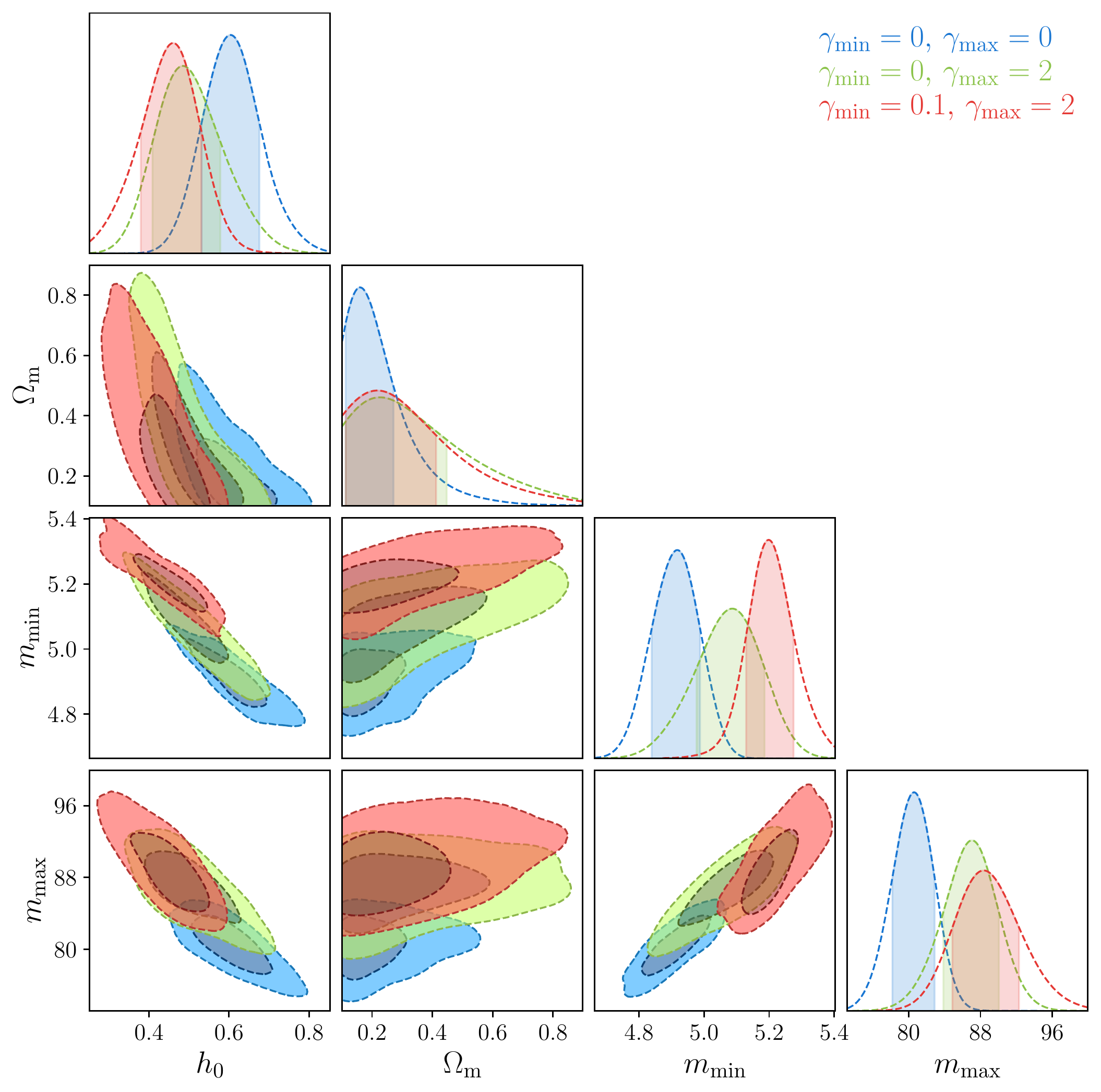}
\caption{Cosmological and population inference with 1,000 2G detections at A+ sensitivity when the fitting model does \emph{not} account for a possible evolution of the mass distribution. Different posterior distributions correspond to different mock catalogs with/without evolution of the minimum and maximum masses. 
Fiducial values are: $h_0=0.677$, $\Om=0.31$, $\mmin=5\Msun$ and $\mmax=78\Msun$.}
\label{fig:corner_plots_noev_all_O5}
\end{figure*}

\begin{figure*}[t!]
\centering
\includegraphics[width = \textwidth,valign=t]{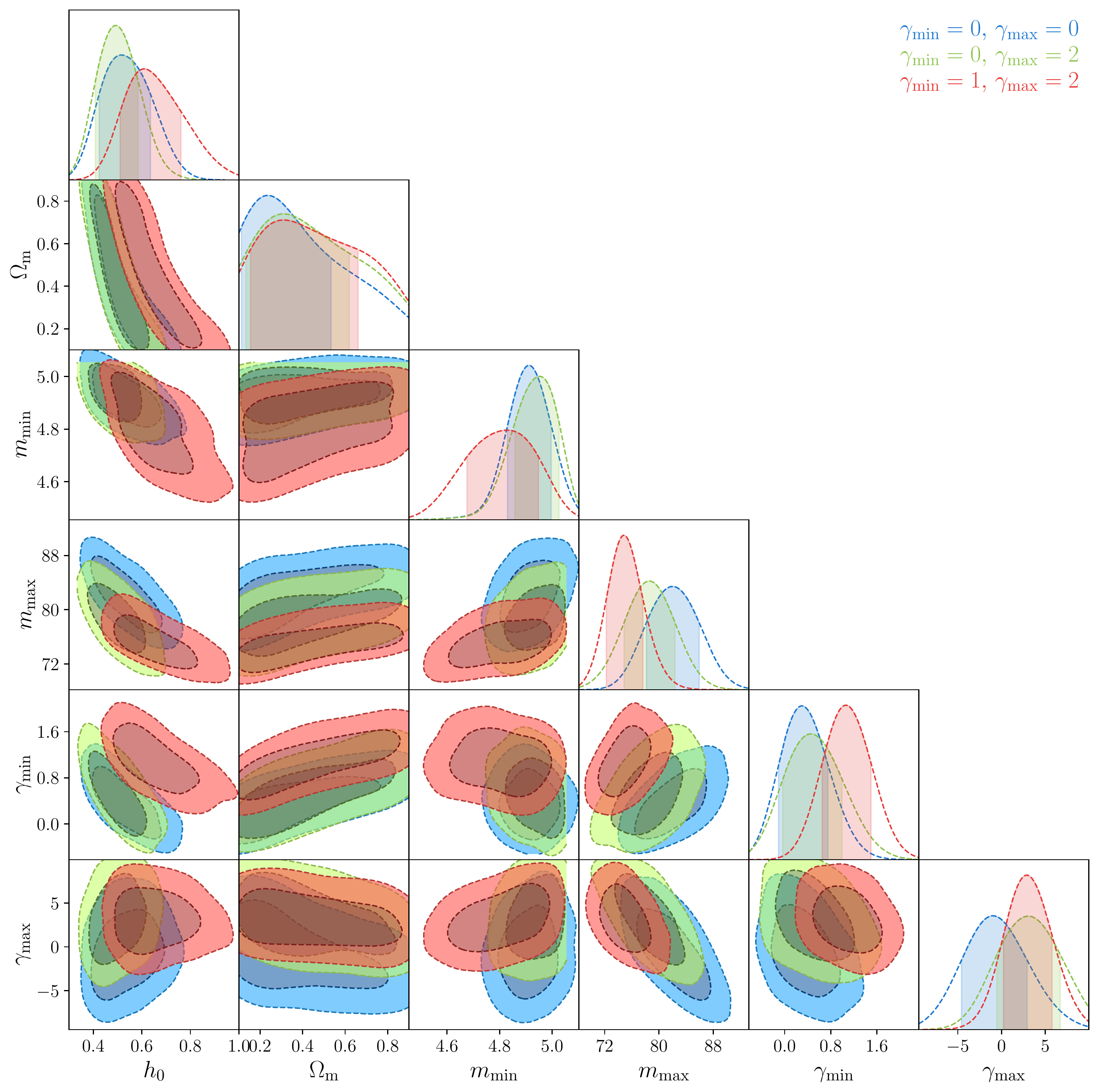}
\caption{Cosmological and population inference with 1,000 2G detections at A+ sensitivity when the fitting model \emph{does} account for a possible evolution of the mass distribution. Different posterior distributions correspond to different mock catalogs as in Fig.~\ref{fig:corner_plots_noev}. 
Fiducial values are: $h_0=0.677$, $\Om=0.31$, $\mmin=5\Msun$, $\mmax=78\Msun$, and the indicated $\gmin$ and $\gmax$.}
\label{fig:corner_plots_O5}
\end{figure*}

\begin{figure*}[t!]
\centering
\includegraphics[width = 0.75\textwidth,valign=t]{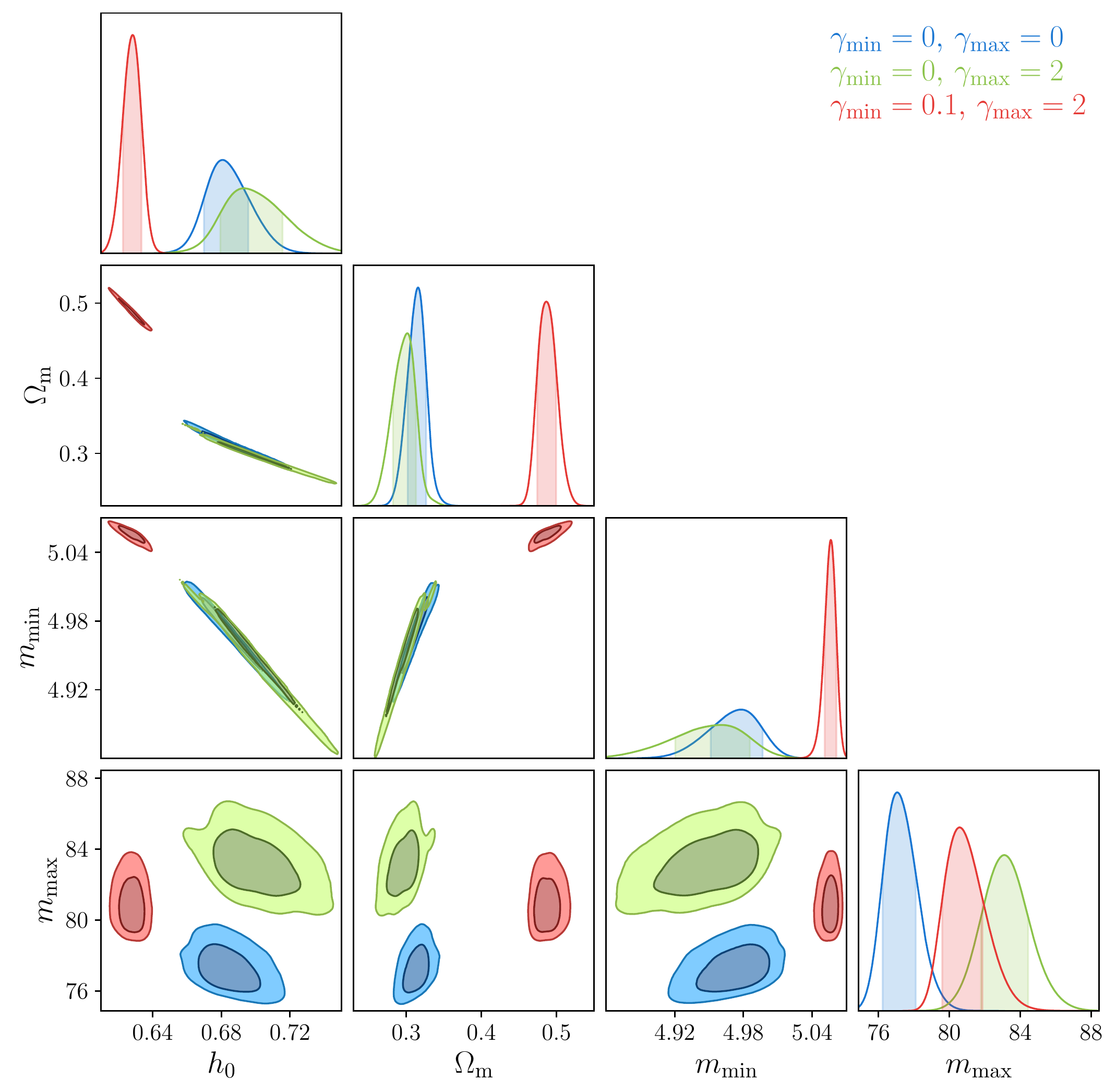}
\caption{Cosmological and population inference with 10,000 3G detections when the fitting model does \emph{not} account for a possible evolution of the mass distribution. Different posterior distributions correspond to different mock catalogs with/without evolution of the minimum and maximum masses.  
Fiducial values are: $h_0=0.677$, $\Om=0.31$, $\mmin=5\Msun$ and $\mmax=78\Msun$. 
The cosmological inference is biased due to the unaccounted redshift evolution of the mass distribution.}
\label{fig:corner_plots_noev_all}
\end{figure*}

\begin{figure*}[t!]
\centering
\includegraphics[width = \textwidth,valign=t]{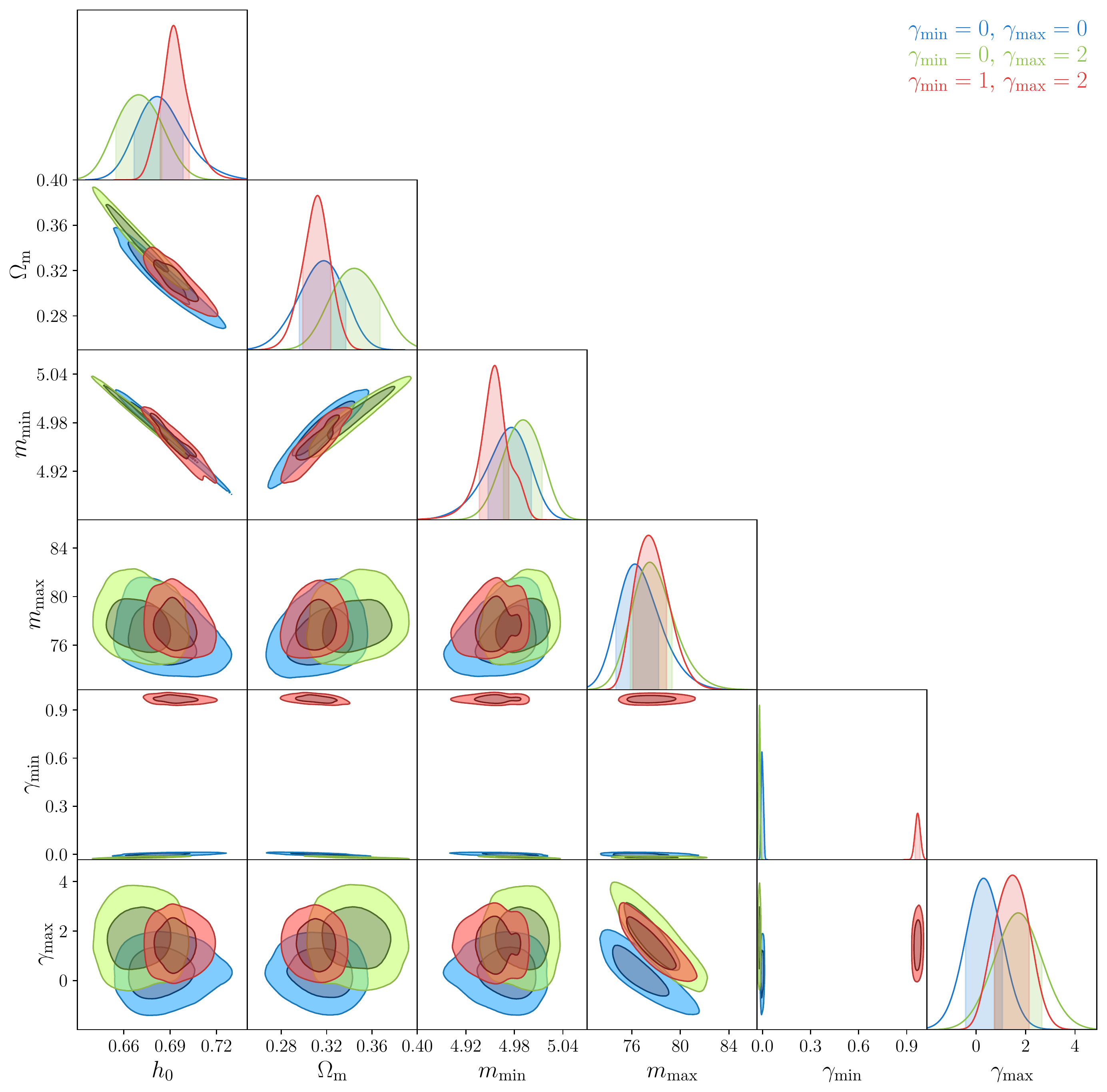}
\caption{Cosmological and population inference with 10,000 3G detections when the fitting model \emph{does} account for a possible evolution of the mass distribution. Different posterior distributions correspond to different mock catalogs with/without evolution of the minimum and maximum masses. 
Fiducial values are: $h_0=0.677$, $\Om=0.31$, $\mmin=5\Msun$, $\mmax=78\Msun$, and the indicated $\gmin$ and $\gmax$.
Both the cosmological and the mass distribution parameters can be recovered.}
\label{fig:corner_plots}
\end{figure*}

\begin{figure*}[t!]
\centering
\includegraphics[width = 0.45\textwidth,valign=t]{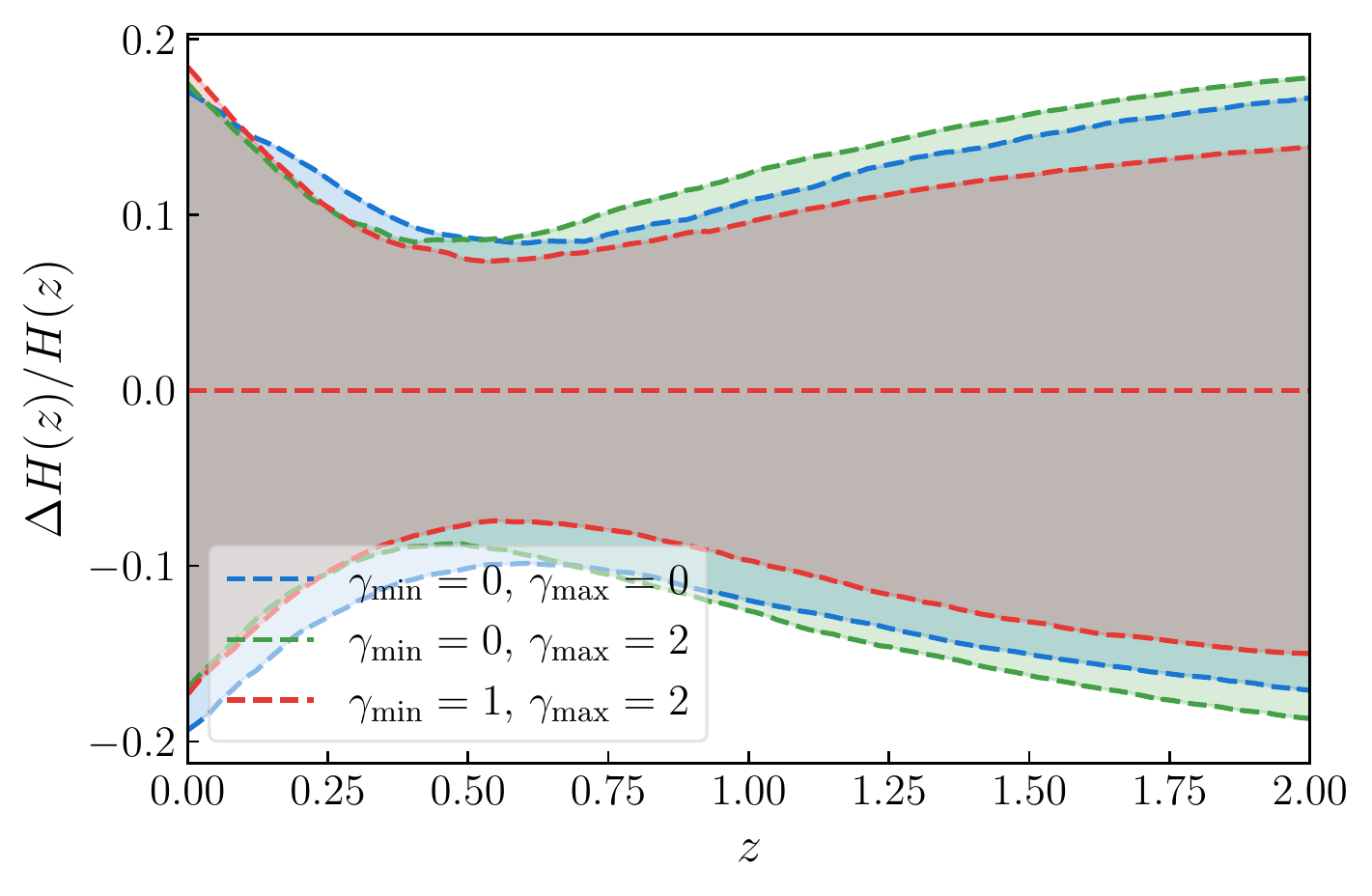}
\includegraphics[width = 0.45\textwidth,valign=t]{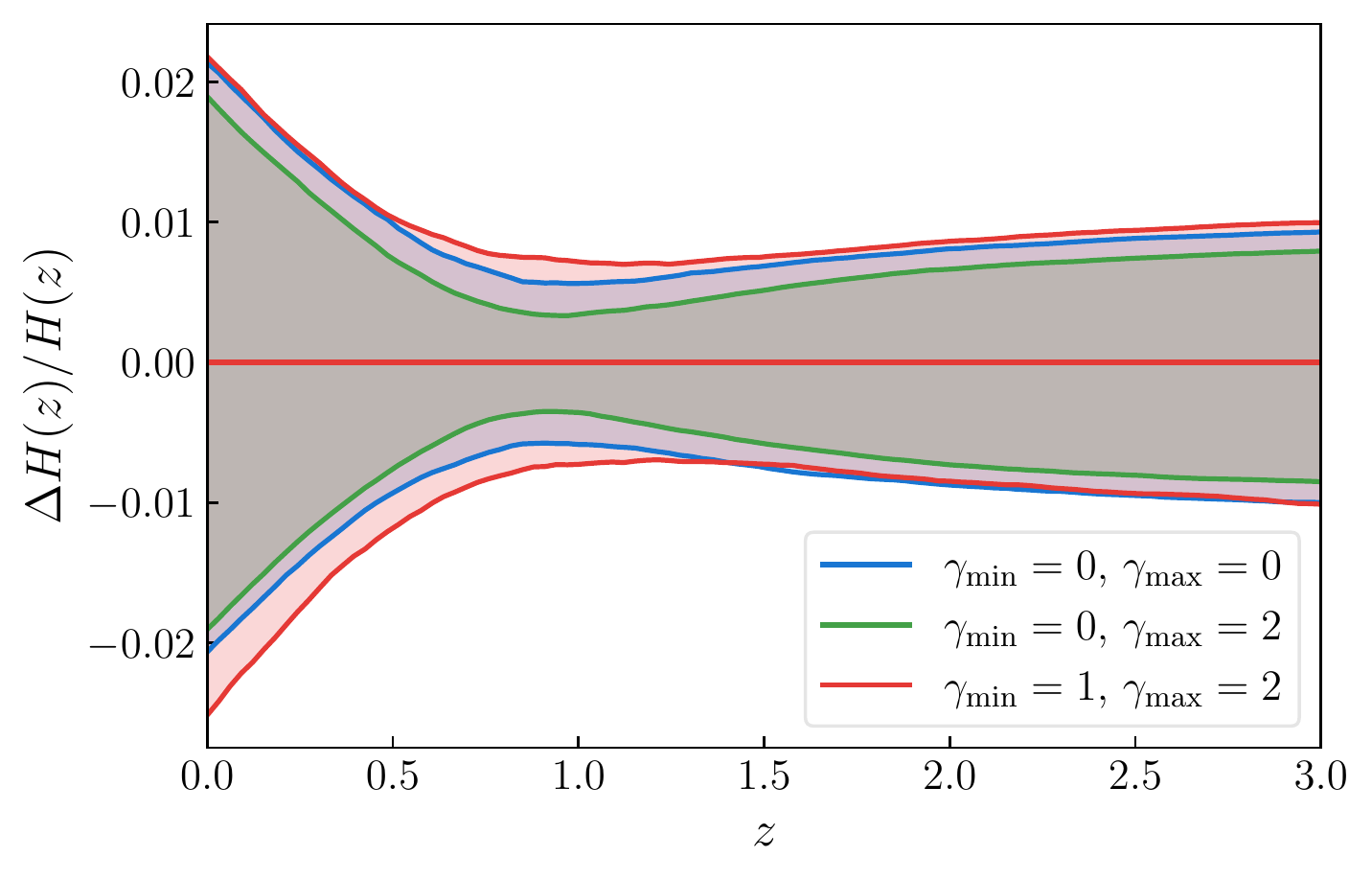}
\caption{$1\sigma$ relative errors in the Hubble parameter with 1,000 2G (left) and 10,000 3G (right) detections when the fitting model \emph{does} account for a possible evolution of the mass distribution. 2G in 1 year could achieve $<10\%$ at $z\sim0.7$, while sub-percent precision is possible with 1 month of observations.}
\label{fig:dHzHz}
\end{figure*}

The posterior distributions of our 2G examples are presented in Fig.~\ref{fig:corner_plots_noev_all_O5} and~\ref{fig:corner_plots_O5}, and for 3G in Fig.~\ref{fig:corner_plots_noev_all} and~\ref{fig:corner_plots}. All figures show three different mock populations with and without evolution in the edges of the mass distribution. 
Figs.~\ref{fig:corner_plots_noev_all_O5} and~\ref{fig:corner_plots_noev_all} correspond to the inference without including the evolution of the mass distribution while Figs.~\ref{fig:corner_plots_O5} and~\ref{fig:corner_plots} include $\gmin$ and $\gmax$ as free parameters. 
We also include the relative errors in $H(z)$ for the different posterior samples in Fig.~\ref{fig:dHzHz}.

We can observe in Fig.~\ref{fig:corner_plots_noev_all_O5} and~\ref{fig:corner_plots_O5} that $\sim1$ year of 2G detectors at A+ sensitivity could provide constraints on the Hubble constant $\sim15\%$ at $1\sigma$ C.L. However, $\Om$ can only be poorly upper bounded $\Om\lesssim0.5$ when there is no evolution and this gets worse when evolution is included. 
The lack of sensitivity to $\Om$ is just a sign of not having enough events at $z\gtrsim1$. 
When there is no evolution in the fitting, Fig.~\ref{fig:corner_plots_noev_all_O5}, $\mmin$ and $\mmax$ are biased in the case of an evolving mock catalog (green and red posteriors). 
When including the evolution parameters in the sampling, Fig.~\ref{fig:corner_plots_O5}, all $\mmin$, $\mmax$, $\gmin$ and $\gmax$ are recovered properly. 
It is to be noted how $\mmin$ and $\mmax$ are both correlated with $h_0$, as the cosmological inference is taking information from both edges in a similar way.

Moving to 3G sensitivities, Fig.~\ref{fig:corner_plots_noev_all} and~\ref{fig:corner_plots}, it is clear the improvement in the determination of all parameters. 
As discussed in the main text, when evolving both edges of the distribution and not taking this into account in the analysis the cosmological inference is highly biased. 
When including the evolution parameters, Fig.~\ref{fig:corner_plots}, this bias disappears. 
Moreover, in comparison with the 2G case, the cosmological parameters $(h_0,\Om)$ are only correlated with $\mmin$, which is the edge that carries most of the weight in the cosmological inference. 

Finally in Fig.~\ref{fig:dHzHz} we plot the relative errors in the Hubble parameter when taking into account the possible mass evolution in the analysis. In both 2G (left) and 3G (right) cases, the errors are similar for all three mock catalogs. 
2G detectors can achieve $\lesssim10\%$ at $z<1$ while 3G maintains $\lesssim1\%$ at $z>1$.

\bibliography{gw_cosmo}
\end{document}